\documentclass[preprint,aps,jmp,nofootinbib]{revtex4}
\usepackage{latexsym}
\usepackage{amsfonts}
\usepackage{amsbsy}
\usepackage{mathrsfs}

\begin{document}

\title{Genuine covariant description of Hamiltonian dynamics}

\date{\today}

\author{Merced Montesinos\footnote{Associate Member of the Abdus Salam
International Centre for Theoretical Physics, Trieste, Italy.}}
\email{merced@fis.cinvestav.mx} \affiliation{Departamento de F\'{\i}sica,
Cinvestav, Av. Instituto Polit\'ecnico Nacional 2508, San Pedro Zacatenco,
07360, Gustavo A. Madero, Ciudad de M\'exico, M\'exico}

\begin{abstract}
After reviewing the covariant description of Hamiltonian dynamics, some
applications are done to the non-relativistic isotropic three-dimensional
harmonic oscillator, Rovelli's model, and $SO(3,1)$ BF theories.
\end{abstract}

\maketitle

\section{Introduction}
In Ref. \cite{Witten} the classical {\it phase space} of a given physical
theory is defined as the space of solutions of its classical equations of
motion. Using this definition, the authors of Ref. \cite{Witten} analyze the
phase spaces of the Klein-Gordon scalar field, Yang-Mills, and general
relativity theories and find that they can be endowed with certain symplectic
structures that come from the handling of the equations of motion of each of
these theories. This fact might leave the impression that the classical
equations of motion of the theory under consideration, in a certain sense,
uniquely determine the symplectic geometry of the phase space. However, such a
conclusion is not correct. In fact, the classical equations of motion do {\it
not} uniquely fix a symplectic structure on the classical phase space even
though its points are in one-to-one correspondence with the solutions of the
classical equations of motion. Thus, the phase space is {\it not} endowed with
a natural or preferred symplectic structure but rather there exists a freedom
in the choice of the symplectic structure on it, which is relevant both
classical and quantum mechanically \cite{MonGer04}. In this sense, a genuine
covariant description of Hamiltonian dynamics consists in choosing different
couples $(\omega,H)$, where $\omega$ is a symplectic structure and $H$ is a
Hamiltonian defined on the same phase space $\Gamma$, for the same dynamical
system. In this paper a brief review of the covariant description of
Hamiltonian systems is performed. The paper contains some new applications,
including gauge systems.

\section{Covariant description of nonparametrized Hamiltonian systems}

\subsection{Canonical viewpoint}
In the standard treatment of Hamiltonian dynamics, the equations of motion are
written in the form \cite{Berndt}
\begin{eqnarray} \label{uhe}
{\dot q}^i = \frac{\partial H}{\partial p_i} \, , \quad {\dot p}_i = -
\frac{\partial H}{\partial q^i} \, , \quad i =1,2, \ldots , n \, ,
\end{eqnarray}
where $H$ is the Hamiltonian of the system, the variables $(q^i, p_i)$ are
canonically conjugate to each other in the sense that
\begin{eqnarray}\label{ccr}
\{ q^i , q^j \} =0 \, , \quad \{ q^i , p_j \} = \delta^i_j \, , \quad \{ p_i ,
p_j \} =0 \, ,
\end{eqnarray}
where $\{ \, , \}$ is the Poisson bracket defined by
\begin{eqnarray}\label{upb}
\{ f , g \} = \frac{\partial f}{\partial q^i} \frac{\partial g}{\partial p_i}
- \frac{\partial f}{\partial p_i} \frac{\partial g}{\partial q^i} \, .
\end{eqnarray}

\subsection{Covariant viewpoint}
The symplectic geometry involved in the Hamiltonian description of mechanics
can clearly be appreciated if Eqs. (\ref{uhe}) are written in the form
\begin{eqnarray}\label{che}
{\dot x}^{\mu} \frac{\partial}{\partial x^{\mu}} = \left ( \omega^{\mu\nu}
\frac{\partial H}{\partial x^{\nu}} \right ) \frac{\partial}{\partial x^{\mu}}
\, ,
\end{eqnarray}
where the independent coordinates $( x^{\mu} ) = ( q^1, \ldots , q^n ; p_1 ,
\ldots , p_n )$ label the points $x$ of the phase space $\Gamma$, $\left \{
\frac{\partial}{\partial x^{\mu}} \right \}$ is a basis of the tangent space
of $\Gamma$ at the point $x$, and
\begin{eqnarray}\label{usm}
\left ( \omega^{\mu\nu} \right ) & = & \left (
\begin{array}{cc}
0 & I \\
- I & 0
\end{array}
\right ) \, ,
\end{eqnarray}
where $0$ and $I$ are $n\times n$ matrices. Moreover, Eq. (\ref{upb}) acquires
the form
\begin{eqnarray}
\{ f , g \}= \frac{\partial f}{\partial x^{\mu}} \omega^{\mu\nu}
\frac{\partial g}{\partial x^{\nu}} \, ,
\end{eqnarray}
from which Eq. (\ref{ccr}) can be rewritten as
\begin{eqnarray}
\{ x^{\mu} , x^{\nu} \} = \omega^{\mu\nu} \, .
\end{eqnarray}
Equations (\ref{che}) are {\it covariant}\/ in the sense that they maintain
their form if the canonical coordinates are replaced by a completely arbitrary
set of coordinates in terms of which $(\omega^{\mu\nu})$ need not be given by
Eq. (\ref{usm}). Similarly, it is possible to retain the original coordinates
$(q^i,p_i)$ and still write the original equations of motion (\ref{uhe}) in
the Hamiltonian form (\ref{che}), but now employing alternative symplectic
structures $\omega^{\mu\nu} (x)$, distinct to that given in Eq. (\ref{usm}),
and taking as Hamiltonian any real function on $\Gamma$ which is a constant of
motion for the system \cite{MonGer04}. This means that the writing of the
equations of motion of a dynamical system in Hamiltonian form is {\it not}
unique \cite{MonGer04,Ger0,Ger3,MonPRA03,Ger,GerSolo,Ger1} (see also Refs.
\cite{Wigner,Hojman}). To better understand the covariant description of
Hamiltonian dynamics, it is convenient to write Eqs. (\ref{che}) in the form
\begin{eqnarray}\label{mmm}
{\dot x}^{\mu} \frac{\partial}{\partial x^{\mu}} = v \, .
\end{eqnarray}
The richness of the covariant description of Hamiltonian systems relies on the
freedom to express the vector field $v$ on the RHS of Eq. (\ref{mmm}) as
$\left ( \omega^{\mu\nu} \frac{\partial H}{\partial x^{\nu}} \right )
\frac{\partial}{\partial x^{\mu}}$ using different couples $(\omega,H)$ and
keeping the same manifold $\Gamma$. Therefore, it is clear that the integral
curves (and the tangent vector field $v$ to them) of any set of equations of
motion are not modified.

The writing of the equations of motion in Hamiltonian form is just half of the
story. The other half concerns with the 1) {\it algebra of observables} and 2)
{\it observables in involution}. An observable ${\cal O}$ is any real-valued
function defined on the phase space $\Gamma$, ${\cal O}: \Gamma
\longrightarrow \mathbb{R}$. If the observable ${\cal O}$ is a constant of
motion, then $d {\cal O}/d t =0$, {\it independently} of the choice of the
symplectic structure $\omega$ and Hamiltonian $H$, which also means that
${\cal O}$ is constant along the vector field $v$
\begin{eqnarray}
0 = \frac{d \cal O}{dt}= \frac{\partial {\cal O}}{\partial x^{\mu}} {\dot
x}^{\mu} = \left ( {\dot x}^{\mu} \frac{\partial}{\partial x^{\mu}} \right )
{\cal O}  = v  {\cal O} \, .
\end{eqnarray}
Nevertheless, even though the analytic form of observables ${\cal O}$ is the
same in any Hamiltonian formulation of a given dynamical system, the algebra
of observables does directly depend on the symplectic structure chosen. As far
as the author knows, this point has been explicitly worked out only in Ref.
\cite{Ger1}. In fact, suppose that the triple $(\Gamma, \omega, H)$ is a
Hamiltonian formulation of a given dynamical system. Suppose that $\{ {\cal
O}_1 , \ldots , {\cal O}_n \}$ is a set of constants of motion which form a
Lie algebra ${\cal A}$ with respect to the symplectic structure $\omega$ of
the triple $(\Gamma, \omega, H)$, $\{{\cal O}_a , {\cal O}_b \}_{\omega}=
c_{ab}\,^c {\cal O}_c$. Now, suppose that the triple $(\Gamma, \Omega, h)$
were another, different from $(\Gamma, \omega, H)$, alternative Hamiltonian
formulation of the {\it same} dynamical system. In the generic case, the
constants of motion of $\{ {\cal O}_1 , \ldots , {\cal O}_n \}$ will not
satisfy $\{ {\cal O}_a , {\cal O}_b \}_{\Omega} =c_{ab}\,^c {\cal O}_c$  with
respect to the symplectic structure $\Omega$ of the triple $(\Gamma, \Omega,
h)$. However, it might be possible to find a new set of constants of motion
$\{ O_1, \ldots , O_n \}$ which might form a new Lie algebra ${\cal A}_{new}$
with respect to the symplectic structure $\Omega$, $\{O_a , O_b \}_{\Omega} =
f_{ab}\,^c O_c$. Therefore, the `algebra of observables' has not an absolute
connotation for a dynamical system. Similarly, if $\{ {\cal I}_1 , \ldots ,
{\cal I}_n \}$ is a set of constants of motion which are in involution with
respect to the symplectic structure $\omega$, i.e, $\{ {\cal I}_i , {\cal I}_j
\}_{\omega} =0$, then the observables of $\{ {\cal I}_1 , \ldots , {\cal I}_n
\}$ are not, in the generic case, in involution with respect to the symplectic
structure $\Omega$. However, it might be possible to find a new set $\{ I_1 ,
\ldots , I_n \}$ which are in involution with respect to the symplectic
structure $\Omega$, $\{ I_i , I_j \}_{\Omega}=0$.

On the other hand, in the same sense that the integral action
\begin{eqnarray}\label{uaction}
S [q^i , p_i ] = \int^{t_2}_{t_1} dt \left [ {\dot q}^i p_i - H(q,p,t) \right
] \, ,
\end{eqnarray}
provides the usual equations of motion (\ref{uhe}), the covariant form of
Hamilton's equations given in Eq. (\ref{che}) can be obtained from the
integral action
\begin{eqnarray}\label{covaction}
S [ x^{\mu} ] = \int^{t_2}_{t_1} d t \left ( \theta_{\mu} (x) \frac{d
x^{\mu}}{dt} - H(x,t)  \right ) \, ,
\end{eqnarray}
provided ${\tilde \delta} S=0$ and ${\tilde \delta} x^{\mu} (t_1)=0=
{\tilde\delta} x^{\mu} (t_2)$ under the arbitrary configurational (or form)
variation of the variables $x^{\mu}$ at $t$ fixed, ${\tilde \delta} x^{\mu}$.
In fact,
\begin{eqnarray}
{\tilde\delta } S & = & \int^{t_2}_{t_1} dt \left ( \omega_{\nu\mu} (x) {\dot
x}^{\mu} - \frac{\partial H}{\partial x^{\nu}} \right ) {\tilde \delta}
x^{\nu} + \left ( \theta_{\mu} (x) {\tilde\delta} x^{\mu} \right )
\mid^{t_2}_{t_1} \, ,
\end{eqnarray}
where $\omega$ is the symplectic two-form, $\omega = \frac12 \omega_{\mu\nu}
(x) d x^{\mu} \wedge d x^{\nu}$ with $\omega_{\mu\nu} = \partial_{\mu}
\theta_{\nu} - \partial_{\nu} \theta_{\mu}$. Equivalently, $\omega= d \theta$
where $\theta= \theta_{\mu} d x^{\mu}$ is the symplectic potential. It is
convenient to make clear some aspects involved with the boundary conditions
${\tilde\delta} x^{\mu} (t_1)=0 = {\tilde\delta} x^{\mu} (t_2)$ employed in
Hamilton's principle. Due to the fact that there are $2n$ coordinates
$x^{\mu}$, one must fix only $2n$ conditions at the time boundary, otherwise
the system might be over-determined, in which case the system might not evolve
from $t_1$ to $t_2$. For instance, in the case when $\theta = p_i d q^i$, it
is clear that one can arbitrarily choose ${\tilde\delta} q^i (t_1)=0$ and
${\tilde \delta} q^i (t_2)=0$. However, in the generic case, namely, when
$\theta= \theta_{\mu} (x) d x^{\mu}$ one can still arbitrarily choose $\delta
x^{\mu} (t_1)=0$ at $t=t_1$. Nevertheless, even though ${\tilde\delta} x^{\mu}
(t_2)=0 $ still holds, $x^{\mu} (t_2)$ cannot be arbitrarily chosen, but it is
fixed by the conditions on $x^{\mu}$ at $t_1$ in order for the system to
evolve from $t_1$ to $t_2$.

Finally, it is important to emphasize that to write a dynamical system in
Hamiltonian form it is not necessary that the symplectic two-form has a
particular symmetry. This constitutes the richness of the Hamiltonian
description of classical dynamics. The relationship between the symmetries of
the action (\ref{covaction}) and constants of motion of the dynamical system
associated to this action is, as usual, through Noether's theorem. In fact, if
under the transformation $t' = t + \delta t $ and ${x'}^{\mu} (t')= x^{\mu}(t)
+ \delta x^{\mu}$ with $\delta = {\tilde\delta} + \delta \frac{d}{dt}$ the
action (\ref{covaction}) transforms as $\delta S= \int^{t_2}_{t_1} \frac{d F
}{dt} dt $ then $\theta_{\mu} \delta x^{\mu} - H \delta t -F$ is a constant of
motion. Further analysis on the implications in the quantum theory of
classical symmetries in the context of the covariant description of
Hamiltonian dynamics can be found in Ref. \cite{otra}.

\subsection{Example: three-dimensional isotropic harmonic oscillator}

The phase space of the system is $\Gamma=\mathbb{R}^6$ whose points can be
labeled with the Cartesian coordinates
$(x^{\mu})=(x^1,x^2,x^3,x^4,x^5,x^6)=(x,y,z,p_x,p_y,p_z)$. The equations of
motion for the system are
\begin{eqnarray}\label{three}
{\dot x} = \frac{p_x}{m}\, ,  \quad {\dot y} = \frac{p_y}{m}\, , \quad {\dot
z}= \frac{p_z}{m}\, , \quad {\dot p}_x = - m \omega^2 x \, , \quad {\dot p}_y
= - m \omega^2 y \, , \quad {\dot p}_z = - m \omega^2 z \, ,
\end{eqnarray}
or, equivalently,
\begin{eqnarray}\label{example}
{\dot x}^{\mu} \frac{\partial}{\partial x^{\mu}}= v \, ,\quad v= \frac{p_x}{m}
\frac{\partial}{\partial x} + \frac{p_y}{m} \frac{\partial}{\partial y} +
\frac{p_z}{m} \frac{\partial}{\partial z} - m \omega^2 \left ( x
\frac{\partial}{\partial p_x} + y \frac{\partial}{\partial p_y} + z
\frac{\partial}{\partial p_z} \right ) \, .
\end{eqnarray}
The canonical viewpoint consists in rewriting the vector field $v$
(\ref{example}) tangent to the integral curves as $v= \left ( \omega^{\mu\nu}
\frac{\partial H}{\partial x^{\nu}} \right ) \frac{\partial}{\partial
x^{\mu}}$ where $\omega^{\mu\nu}$ is that of Eq. (\ref{usm}) and $H$ is the
energy for the system, $H= \frac{1}{2m} \left ( (p_x)^2 + (p_y)^2 + (p_z)^2
\right ) + \frac12 m \omega^2 \left ( x^2 + y^2 + z^2 \right )$. However,
according to the covariant viewpoint of Hamiltonian dynamics, it is possible
to give alternative descriptions of the dynamics of the system, one of which,
for instance, consists in rewriting the vector field $v$ of Eq.
(\ref{example}) as $v= \left ( \omega^{\mu\nu} \frac{\partial H}{\partial
x^{\nu}} \right ) \frac{\partial}{\partial x^{\mu}}$ with
\begin{eqnarray}
\left ( \omega^{\mu\nu} \right ) & = & \left (
\begin{array}{rrrrrr}
0 & 0 & 0 & 0 & 0 & 1\\
0 & 0 & 0 & 0 & 1 & 0 \\
0 & 0 & 0 & 1 & 0 & 0 \\
0 & 0 & -1 & 0 & 0 & 0 \\
0 & -1 & 0 & 0 & 0 & 0 \\
-1 & 0 & 0 & 0 & 0 & 0
\end{array}
\right ) \, ,\quad H = \frac{p_x p_z}{m} + m \omega^2 x z + \frac{(p_y)^2}{2m}
+ \frac12 m \omega^2 y^2 \, ,
\end{eqnarray}
among many other possibilities.

\section{Covariant description of gauge systems}
Even though the covariant description of nonparametrized Hamiltonian systems
has been analyzed, the covariant description of gauge systems has not been
systematically studied. As far as the author knows, the first paper dealing
with the covariant Hamiltonian description, in the sense of the present paper,
of constrained systems is Ref. \cite{MonMon}, where parametrized
nonrelativistic Hamiltonian systems were analyzed in detail and where also the
first steps towards the covariant Hamiltonian description of arbitrary gauge
systems (in the context of constraints) were done. So, we continue the
analysis of the generic case of gauge systems. Moreover, we make clear some
aspects of gauge systems that were not clearly stated in such a paper. Also,
some minor corrections to that part of the paper are done here. After all, one
has in mind realistic theories, so it is relevant to understand how the
covariant formulation of gauge systems works. This is the issue of this
section.

\subsection{Canonical viewpoint}
Suppose one has a dynamical system defined by the dynamical equations of
motion
\begin{eqnarray}\label{eqmotion}
{\dot q}^i & = & \lambda^a \frac{\partial \gamma_a}{\partial p_i} \, , \quad
{\dot p}_i  =  - \lambda^a \frac{\partial \gamma_a}{\partial q^i} \, ,
\end{eqnarray}
where $\cdot = \frac{d}{d \tau}$ and subject to the constraints
\begin{eqnarray}\label{cs}
\gamma_a \approx 0 \, ,
\end{eqnarray}
which satisfy
\begin{eqnarray}
\{ \gamma_a , \gamma_b \} = C_{ab}\,^c \gamma_c \, ,
\end{eqnarray}
with respect to the canonical symplectic structure of Eq. (\ref{upb}), and so
the $\gamma$'s are first class \cite{Dirac}. If the coordinates $q$'s and
$p$'s which label the points of the extended phase space $\Gamma_{ext}$ are
grouped in the new variable $x^{\mu}$, $\mu=1,\ldots,2n$ such that $(x^1, x^2
, \ldots, x^n, x^{n+1}, x^{n+2},\ldots, x^{2n})=(q^1, q^2, \ldots, q^n, p_1,
p_2, \ldots, p_n)$ then the dynamical equations of Eq. (\ref{eqmotion})
acquire the form
\begin{eqnarray}\label{eqmotionII}
{\dot x}^{\mu} \frac{\partial}{\partial x^{\mu}}  = \lambda^a X_a \, ,
\end{eqnarray}
where the vector fields $X_a$ are given by
\begin{eqnarray}\label{generators}
X_a = X^{\mu}_a  \frac{\partial}{\partial x^{\mu}} = \left ( \omega^{\mu\nu}
\frac{\partial \gamma_a}{\partial x^{\nu}} \right ) \frac{\partial}{\partial
x^{\mu}} \, ,
\end{eqnarray}
with $(\omega^{\mu\nu})$ given by Eq. (\ref{usm}).

\subsection{Covariant viewpoint}
There exists also the possibility of making a different choice of both the
symplectic structure $\omega$ and the analytical form of the constraints
$\gamma_a$ in Eqs. (\ref{eqmotionII}) and (\ref{generators}). Let $\Omega$ and
$G_a$ be a new symplectic structure and a different analytical form of the
constraints; respectively. It is important to emphasize that the extended
phase space $\Gamma_{ext}$ as a manifold is the {\it same}\/ as the one
before. In particular, the points ${\bf p}\in \Gamma_{ext}$ can still be
labeled with the same coordinates $x^{\mu}$ as before. Of course, one can
choose another set of coordinates, but this is not relevant for the present
discussion and the introduction of a different set of coordinates is avoided
just to emphasize that the alternative symplectic structures and the form of
the constraints have a different analytical form with respect to the ones of
the standard viewpoint. In summary, $\Omega \neq \omega$ and $G_a \neq
\gamma_a$ on $\Gamma_{ext}$, which is a statement which holds independently of
the particular set of coordinates chosen. Now, the idea is to express the
vector fields $X_a$ of Eq. (\ref{generators}) in terms of the new symplectic
structure $\Omega$ and the new constraints $G_a$ in such a way that the vector
fields $X_a$ remain the {\it same}, i.e.,
\begin{eqnarray}
X_a = X^{\mu}_a  \frac{\partial}{\partial x^{\mu}} = \left ( \Omega^{\mu\nu}
\frac{\partial G_a}{\partial x^{\nu}} \right ) \frac{\partial}{\partial
x^{\mu}} \, .
\end{eqnarray}
What happens to the dynamical equations of motion given in Eq.
(\ref{eqmotion})? These equations of motion are the same ones than those given
in Eq. (\ref{eqmotionII}). Therefore, due to the fact that the vector fields
$X_a$ are the same, the dynamical equations are {\it not} modified, i.e., they
are also the same in the present case. What about the constraint surface
$\Sigma$? Well, in the present case the constraint surface $\Sigma$ is defined
by
\begin{eqnarray}
G_a \approx 0 \, .
\end{eqnarray}
Thus, it is obvious that the analytical form of the equations that define the
constraint surface of this case are completely different to those equations of
the starting point, given in Eq. (\ref{cs}). It is also important to emphasize
that new constraints $G_a$ are {\it not}, in the generic case, a linear
combination of the original set of constraints $\gamma_a$. So, the freedom in
the choice of the constraint surface has nothing to do with the usual
abelianization of the constraints $\gamma_a$ simply because the abelianization
procedure is not accompanied with a change in the symplectic structure as it
happens in the present context. This fact is clearly appreciated even in the
case of parametrized nonrelativistic Hamiltonian systems \cite{MonMon}. So,
the new constraints $G_a$ are first class with respect to the new symplectic
structure $\Omega$. An action principle, which gives the covariant form of the
equations of motion is
\begin{eqnarray}
S[x^{\mu}, \lambda^a ] & = & \int^{\tau_2}_{\tau_1} d \tau \left [
\theta_{\mu} (x) {\dot x}^{\mu} - \lambda^a \gamma_a (x) \right ] \, ,
\end{eqnarray}
with $\omega_{\mu\nu}= \partial_{\mu} \theta_{\nu} - \partial_{\nu}
\theta_{\mu}$ and the constraints $\gamma_a$ are first class with respect to
$\omega=\frac12 \omega_{\mu\nu} d x^{\mu} \wedge d x^{\nu}$.

\subsection{Examples}

\subsubsection{Rovelli's model.}The extended phase space of this model is
$\Gamma_{ext}=\mathbb{R}^4$. The points $p\in \Gamma_{ext}$ are labeled with
the cartesian coordinates $(x^{\mu})=(q^1,q^2,p_1,p_2)$. In these coordinates,
the symplectic structure $\omega$ on $\Gamma_{ext}$ is defined by
\begin{eqnarray}
\omega = d p_1 \wedge d q^1 + dp_2 \wedge d q^2 \, .
\end{eqnarray}
The dynamical equations are
\begin{eqnarray}
{\dot q}^1 & = & \lambda p_1 \, , \quad {\dot q}^2 =  \lambda p_2 \, , \quad
{\dot p}_1 = - \lambda q^1 \, , \quad {\dot p}_2 = - \lambda q^2 \, .
\end{eqnarray}
where $\lambda$ is a Lagrange multiplier \cite{RovBook}. On the other hand,
the constraint surface $\Sigma$ is
\begin{eqnarray}
\gamma := \frac12 \left ( (p_1)^2 + (p_2)^2 + (q^1)^2 + (q^2)^2 \right ) -M
\approx 0 \, ,
\end{eqnarray}
where $M$ is a positive constant. From the dynamical equations, the vector
field $X$ is read off
\begin{eqnarray}
X = p_1 \frac{\partial}{\partial q^1} +  p_2 \frac{\partial}{\partial q^2} -
q^1 \frac{\partial}{\partial p_1} - q^2 \frac{\partial}{\partial p_2} = \left
( \omega^{\mu\nu} \frac{\partial \gamma}{\partial x^{\nu}} \right )
\frac{\partial}{\partial x^{\mu}}
\end{eqnarray}

Using the discussion of the preceding subsection as well as Ref.
\cite{MonMon}, Rovelli's model can, alternatively, be described, for instance,
as follows:

1) The extended phase space remains the same, $\Gamma_{ext}=\mathbb{R}^4$. If
the symplectic structure on $\Gamma_{ext}$ is chosen to be
\begin{eqnarray}
\omega_1 & = & d p_2 \wedge d q^1 + d p_1 \wedge d q^2 \, ,
\end{eqnarray}
then the new constraint surface is defined by
\begin{eqnarray}
\gamma_1 := \left ( p_1 p_2 + q^1 q^2 \right ) - N \approx 0 \, ,
\end{eqnarray}
where $N$ is a constant. The vector field $X$ is rewritten as
\begin{eqnarray}
X = p_1 \frac{\partial}{\partial q^1} +  p_2 \frac{\partial}{\partial q^2} -
q^1 \frac{\partial}{\partial p_1} - q^2 \frac{\partial}{\partial p_2} = \left
( \omega^{\mu\nu}_1 \frac{\partial \gamma_1}{\partial x^{\nu}} \right )
\frac{\partial}{\partial x^{\mu}} \, .
\end{eqnarray}

2) The extended phase space remains the same, $\Gamma_{ext}=\mathbb{R}^4$. If
the symplectic structure on $\Gamma_{ext}$ is chosen to be
\begin{eqnarray}
\omega_2 & = & - d p_1 \wedge d q^1 + d p_2 \wedge d q^2 \, ,
\end{eqnarray}
then the new constraint surface is defined by
\begin{eqnarray}
\gamma_2 := \frac12 \left ( (p_2)^2 - (p_1)^2 + (q^2)^2 - (q^1)^2 \right ) - L
\approx 0 \, ,
\end{eqnarray}
where $L$ is a constant. The vector field $X$ is rewritten as
\begin{eqnarray}
X = p_1 \frac{\partial}{\partial q^1} +  p_2 \frac{\partial}{\partial q^2} -
q^1 \frac{\partial}{\partial p_1} - q^2 \frac{\partial}{\partial p_2} = \left
( \omega^{\mu\nu}_2 \frac{\partial \gamma_2}{\partial x^{\nu}} \right )
\frac{\partial}{\partial x^{\mu}} \, .
\end{eqnarray}

3) The extended phase space remains the same, $\Gamma_{ext}=\mathbb{R}^4$. If
the symplectic structure on $\Gamma_{ext}$ is chosen to be
\begin{eqnarray}
\omega_3 & = & d q^1 \wedge d q^2 + d p_1 \wedge d p_2 \, ,
\end{eqnarray}
then the new constraint surface is defined by
\begin{eqnarray}
\gamma_3 := \left ( q^1 p_2 - q^2 p_1 \right ) - R \approx 0 \, ,
\end{eqnarray}
where $R$ is a constant. The vector field $X$ is rewritten as
\begin{eqnarray}
X = p_1 \frac{\partial}{\partial q^1} +  p_2 \frac{\partial}{\partial q^2} -
q^1 \frac{\partial}{\partial p_1} - q^2 \frac{\partial}{\partial p_2} = \left
( \omega^{\mu\nu}_3 \frac{\partial \gamma_3}{\partial x^{\nu}} \right )
\frac{\partial}{\partial x^{\mu}} \, .
\end{eqnarray}

\section{Field theory}
In this section, an application to field theory is done. The system is defined
by the equations of motion
\begin{eqnarray}\label{BFem}
F^I\,_J [A] & = & 0 \, , \quad D B^{IJ} = 0 \, ,
\end{eqnarray}
where $F^I\,_J [A] = d A^I\,_J + A^I\,_K \wedge A^K\,_J$ is the curvature of
the Lorentz connection $A^I\,_J =A_{\mu}\,^I\,_J d x^{\mu}$ and $DB^{IJ}=d
B^{IJ} + A^I\,_K \wedge B^{KJ} + A^J\,_K \wedge B^{IK}$ where $B^{IJ}$ is a
set of six two-forms on account of $B^{IJ}=-B^{JI}$. The equations of motion
(\ref{BFem}) are usually obtained from the BF action
\begin{eqnarray}\label{BFaction}
S [A,B] & = & a_1 \int_{\mathscr{M}} B^{IJ} \wedge F_{IJ} [A] \, ,
\end{eqnarray}
where the Lorentz indices $I,J,\ldots=0,1,2,3$ are raised and lowered with the
Minkowski metric $\eta_{IJ}$, $(\eta_{IJ})=\mbox{diag} (-1,+1,+1,+1)$.
Defining $B^i =- \frac12 \varepsilon^i\,_{jk} B^{jk}$ and $\Gamma^i = -\frac12
\varepsilon^i\,_{jk} A^{jk}$ where the Latin indices $i,j, \ldots$ are raised
and lowered with the Euclidean metric $\delta_{ij}$,
$(\delta_{ij})=\mbox{diag} (+1,+1,+1)$, the Hamiltonian analysis of the action
(\ref{BFaction}) leads to
\begin{eqnarray}
S & = & \int d x^0 \int_{\Sigma} d^3 x \left [ {\dot \Gamma}_a\,^i {\widetilde
\pi}^a\,_i + {\dot A}_a\,^{0i} {\widetilde p}^a\,_i - \lambda^i {\widetilde
G}_i - \Lambda^i {\widetilde H}_i - \lambda_{ai} {\widetilde C}^{ai} -
\rho_{ai} {\widetilde D}^{ai} \right ]\, ,
\end{eqnarray}
where it has been assumed that $\mathscr{M}$ has the topology $\Sigma\times
R$, with $\Sigma$ closed (compact and without boundary) to avoid boundary
terms. The four-dimensional coordinates $(x^{\mu})=(x^0,x^a)$, $a,b,\ldots
=1,2,3$ are such that $x^a$ label points on $\Sigma$. The totally
anti-symmetric Levi-Civita density of weight $+1$, ${\widetilde
\eta}^{\alpha\beta\mu\nu}$, is such that ${\widetilde \eta}^{0123}=+1$. Also
${\widetilde \eta}^{0abc}={\widetilde \eta}^{abc}$. The other definitions
involved are ${\widetilde \pi}^a\,_i := a_1 {\widetilde \eta}^{abc}
B_{bc\,\,i}$, ${\widetilde p}^a\,_i := a_1 {\widetilde \eta}^{abc}
B_{bc\,\,0i}$, $\lambda^i := -\Gamma_0\,^i$, $\Lambda^i := - A_{0}\,^{0i}$,
$\lambda_{ai}:= -2 a_1 B_{0a\,\,i}$, $\rho_{ai}:= - 2 a_1 B_{0a\,\,0i}$, and
\begin{eqnarray}\label{const}
{\widetilde G}_i & := & \partial_a {\widetilde \pi}^a\,_i +
\varepsilon_{ij}\,^k \Gamma_a\,^j {\widetilde \pi}^a\,_k +
\varepsilon_{ij}\,^k A_a\,^{0j} {\widetilde p}^a\,_k \, , \nonumber\\
{\widetilde H}_i & = & \partial_a {\widetilde p}^a\,_i + \varepsilon_{ij}\,^k
\Gamma_a\,^j {\widetilde p}^a\,_k -
\varepsilon_{ij}\,^k A_a\,^{0j} {\widetilde \pi}^a\,_k \, , \nonumber\\
{\widetilde C}^{ai} & := & \frac12 {\widetilde \eta}^{abc} F_{bc}\,^i \, ,
\nonumber\\
{\widetilde D}^{ai} & := & \frac12 {\widetilde \eta}^{abc} F_{bc}\,^{0i} \, ,
\end{eqnarray}
where $F_{bc}\,^i := - \frac12 \varepsilon^i\,_{jk} F_{bc}\,^{jk}$. Finally,
the symplectic structure is given by (see also Ref. \cite{mo})
\begin{eqnarray}\label{se}
\{ \Gamma_a\,^i (x^0,x) , {\widetilde \pi}^b\,_j (x^0 ,y) = \delta^b_a
\delta^i_j \delta^3 (x,y)\, , \quad \{ A_a\,^{0i} (x^0,x) , {\widetilde
p}^b\,_j (x^0,y) \} = \delta^b_a \delta^i_j \delta^3 (x,y) \, .
\end{eqnarray}

Alternatively, it is also possible to take the action principle
\begin{eqnarray}\label{newBFaction}
S_2 [A,B] & = & a_2 \int_{\mathscr{M}} \ast B^{IJ} \wedge F_{IJ} [A] \, ,
\end{eqnarray}
with $\ast B^{IJ}=\frac12 \varepsilon^{IJ}\,_{KL} B^{KL}$ and
$\varepsilon_{0123}=+1$. The variation of the action (\ref{newBFaction}) with
respect to the connection $A^I\,_J$ and the $B^{IJ}$ fields yields $D \ast
B^{IJ}=0$ and $\ast F^{IJ}=0$, which after the application of the dual
operation ``$\ast$" reduce to those given in Eq. (\ref{BFem}). So, both
actions (\ref{BFaction}) and (\ref{newBFaction}) give rise to the same
equations of motion. In spite of this, the symplectic structures defined by
both actions are different from each other. In fact, the Hamiltonian analysis
of the action (\ref{newBFaction}) leads to
\begin{eqnarray}
S_2 = \int d x^0 \int_{\Sigma} \left [ \frac{a_2}{a_1} {\dot \Gamma}_a\,^i
{\widetilde p}^a\,_i  - \frac{a_2}{a_1} {\dot A}_a\,^{0i}
{\widetilde\pi}^a\,_i + \frac{a_2}{a_1} \Lambda^i {\widetilde G}_i -
\frac{a_2}{a_1} \lambda^i {\widetilde H}_i - \frac{a_2}{a_1} \rho_{ai}
{\widetilde C}^{ai} + \frac{a_2}{a_1} \lambda_{ai} {\widetilde D}^{ai} \right
] \, .
\end{eqnarray}
The symplectic structure in this case is given by
\begin{eqnarray}
\{ \Gamma_a\,^i (x^0,x) , \frac{a_2}{a_1} {\widetilde p}^b\,_j (x^0,y) \} =
\delta^b_a \delta^i_j \delta^3 (x,y) \, , \quad \{ A_a\,^{0i} (x^0,x) , -
\frac{a_2}{a_1} {\widetilde \pi}^b\,_j (x^0,y) \} = \delta^b_a \delta^i_j
\delta^3 (x,y) \, .
\end{eqnarray}
[cf. Eq. (\ref{se})].

\section*{Acknowledgements}
The author wishes to thank G.F. Torres del Castillo for very fruitful
discussions. This work was supported in part by the CONACyT grant no.
SEP-2003-C02-43939.


\end{document}